\documentstyle[sprocl]{article}

\input{psfig}

\def\Journal#1#2#3#4{{#1} {\bf #2}, #3 (#4)}

\def\PRD{{\em Phys. Rev.} D}

\begin{document}

\begin{titlepage}
\mbox{ }
\vspace{-2.0 cm}

\begin{tabbing}
xxxxxxxxxxxxxxxxxxxxxxxxxxxxxxxxxxxxxxxxxxxxxxxxxxxx\=xxxxxxxxxx\kill
 \> IUHET-392 \\ 
 \> \hfill July 1998 \\
\end{tabbing}

\vspace{1.5 cm}

\begin{center}

\renewcommand{\thefootnote}{\fnsymbol{footnote}}

{\Large\bf MSSM with Large tan$\beta$ Constrained
       by Minimal SO(10) Unification}~\footnote
{Talk presented at PASCOS-98, Boston, MA, March 1998.
 To appear in the proceedings.}

\vspace{1.0 cm}

{\large\bf  Tom\'{a}\v{s} Bla\v{z}ek~\footnote
            {On leave of absence from the Dept.\ of Theoretical Physics, 
            Comenius Univ., Bratislava, Slovakia; current e-mail address: 
            {\em blazek@indiana.edu}}}\\ 
{\em Department of Physics, Indiana University, Swain Hall West 117, Bloomington, IN 47405, USA}

\vspace{2.0 cm}

%\vspace{2,3 cm}

{\bf Abstract}

\end{center}

Study of the MSSM in large tan$\beta$ regime has to include correlations 
between the constraints presented by the low energy values of the b quark 
mass and $BR(b\rightarrow s\gamma)$. Both quantities receive SUSY 
contributions  
enhanced by tan$\beta$ and have a major impact on the MSSM analysis. Here 
we summarize the results of such a study constrained by minimal SO(10) 
unification. We show the best fits, in the $(m_0,M_{1/2})$ 
plane, obtained in the global analysis spanned over the gauge, Yukawa and
SUSY parameter space at the unification scale. Two distinct fits 
describe the available low-energy data very well. The fits differ by the 
overall sign of the $b\rightarrow s\gamma$ decay amplitude. 
We conclude that an attractive SO(10)-derived regime of the MSSM remains
a viable option.

\vfill

\setcounter{footnote}{0}
\renewcommand{\thefootnote}{\arabic{footnote}}

%\noindent PACS numbers: 12.15.Ff, 12.15.Hh, 12.60.Jv
\end{titlepage}

%
% end of the preprint page
%
%%%%%%%%%%%%%%%%%%%%%%%%%%%%%%%%%%%%%%%%%%%%%%%%%%%%%%%%%%%%%%%%%%%%%%%%
%%BEGINNING OF TEXT                           
%%%%%%%%%%%%%%%%%%%%%%%%%%%%%%%%%%%%%%%%%%%%%%%%%%%%%%%%%%%%%%%%%%%%%%%%

\title{MSSM WITH LARGE tan$\beta$ CONSTRAINED
       BY MINIMAL SO(10) UNIFICATION}

\author{T. BLA\v{Z}EK}

\address{Department of Physics, Indiana University,\\
         Swain Hall West 117, Bloomington, IN 47405, USA\\
         E-mail: blazek@indiana.edu}

\maketitle\abstracts{ 
Study of the MSSM in large tan$\beta$ regime has to include correlations 
between the constraints presented by the low energy values of the b quark 
mass and $BR(b\rightarrow  s \gamma)$. Both quantities receive SUSY 
contributions 
enhanced by tan$\beta$ and have a major impact on the MSSM analysis. Here 
we summarize the results of such a study constrained by minimal SO(10) 
unification. We show the best fits, in the $(m_0,M_{1/2})$ 
plane, obtained in the global analysis spanned over the gauge, Yukawa and
SUSY parameter space at the unification scale. Two distinct fits 
describe the available low-energy data very well. The fits differ by the 
overall sign of the $b\rightarrow s \gamma$ decay amplitude. 
We conclude that an attractive SO(10)-derived regime of the MSSM remains
a viable option.}

\section{Motivation}
Detailed global analysis of the unconstrained Minimal Supersymmetric Standard
Model (MSSM) is of utmost importance.
However, the MSSM is understood as
just an effective description of more fundamental interactions.
Pursuing a search for such a more fundamental theory, global
analysis can then also be used to evaluate different
models which break down to the MSSM at some high scale. In this sense, our
work was motivated by simple SO(10) grand unified theories\cite{adhrs}
(GUTs). Amazingly, these can be made simple enough 
to have positive number of degrees of freedom and thus yield a 
sensible analysis. Note that simplicity of these models implies large
tan$\beta$, and that in turn requires to include the supersymmetric (SUSY)
sector into the analysis because of potentially significant SUSY
threshold corrections. To simplify the analysis we assumed universal
sparticle masses (with the exception of the non-universal scalar Higgs
masses) and scalar couplings emerging at the GUT scale. 
That leaves flavor physics (the SO(10)-derived fermion mass matrices) 
as the main issue of the analysis. The point of this talk is not, however,
the origin of flavor. The point here is 
that once a candidate model which fits the observed fermion masses and 
mixings very well is found, an SO(10) global analysis resumes the role
of the MSSM global analysis constrained by unification. In fact, it is 
only the fermionic sector which substantially distinguishes the two.

In particular, it was shown in ref.\cite{bcrw} that model 4c 
is an example of such a nice model. When $\chi^2<2\; per\;d.o.f.$,
only 20-50\% of the total $\chi^2$ comes from the masses of the
lighter-generation fermions and fermion mixings while the remaining
$\chi^2$ contributions originate from the observables like gauge boson
masses\footnote 
{
In our global analysis,
we assume a pure top-down
approach with $\mu$ and $B\mu$
as initial parameters instead of
the $Z$ boson mass, pseudoscalar 
mass $m_{A^0}$, or tan$\beta$
used in bottom-up approach.
The conditions for 1-loop radiative
electroweak symmetry breaking can then
result in inaccurate gauge boson masses
which are brought down to the agreement 
with the data only in the course of 
optimization.
},
gauge couplings, third-generation fermion masses, and 
$BR(b\rightarrow s\gamma)$ --- the observables 
traditionally included in the MSSM analysis constrained by unification.
For this reason the features of the model 4c best fits are the features
which come out of the MSSM analysis with unification constraints. 

\section{Results}
The results indicate that there are two major constraints imposed on the
MSSM with large tan$\beta$: the observed $b\rightarrow s\gamma$
decay rate and the value of the bottom quark mass $m_b$. Both get a SUSY
contribution enhanced by large tan$\beta$. When considered separately in
the SUSY parameter space, each one tends to prefer a different sign of the 
$\mu$ parameter\footnote
{
In fact, we should talk about the sign of $\mu A_t$, but at the electroweak
scale the stop trilinear coupling always turns out negative in acceptable
fits. 
}. 
However, they can both be satisfied by the sign of $\mu$ favored
by $b\rightarrow s\gamma$. Fits with the opposite sign of $\mu$ can
reproduce the observed $BR(b\rightarrow s\gamma)$ only for the SUSY
spectrum deep in the TeV region. The correct sign of $\mu$ (positive in our
notation) means a destructive interference among the chargino and 
(SM + charged Higgs) contributions to $b\rightarrow s\gamma$, and positive
SUSY threshold correction to $m_b$. Both have grave consequences.

The destructive interference among the $b\rightarrow s\gamma$ amplitudes
can result in having the net amplitude $C_7^{MSSM}$ of the same sign as 
$C_7^{SM}$, or of the opposite sign. The opposite sign does not look too
unnatural --- after all, the chargino contribution is enhanced by
tan$\beta$ (and suppressed only by sparticle masses) compared to the SM
contribution\footnote
{
That also explains why the other sign of $\mu$ is so strongly constrained.
}.
Hence two distinct fits are possible (figs.1a-b). 
Note that the fit with the flipped sign (fig.1b) allows for lighter SUSY
spectrum to make the destructive interference work. For the same reason the
charged Higgs (and then the rest of the Higgs sector) tends to be heavier
when the sign is flipped and lighter (in the best fits, as light as
experimentally allowed) in the fit of fig.1a.  
The detailed analysis\cite{br} of $C_7^{MSSM}$ shows that one cannot
neglect the contribution from the inter-generational squark mixing as is
sometimes assumed. However, 
this term makes the analysis dependent on the choice of a GUT model (in 
addition to the pattern of SUSY breaking).

The $b$ quark mass is a tight constraint in this regime. The best fits
make $\alpha_s(M_Z)$ low 
to minimize the renormalization of $m_b$ at low energies.
$\alpha_s(M_Z)\le 0.118$ is achieved by a negative 3-5\% GUT threshold 
correction to $\alpha_s$ coming from mass splits in superheavy
multiplets. SUSY parameters are also adjusted to prevent too large
$\delta m_b^{SUSY}>0$. Most importantly, 
the magnitude of $\mu$ is obtained as low as allowed by
sparticle searches. Thus the lightest possible higgsino-like
chargino and neutralino are present in the best fits (but are not 
necessary: good fits --- shifted towards greater $m_0,\,M_{1/2}$ as the
price to pay --- survive a further increase of the experimental gaugino
mass limits). 

The best fit value of $A_t(M_Z)$ 
varies across the $(m_0,M_{1/2})$ plane by about 1000GeV, 
(and universal coupling $A_0$ varies even more)  
since it is a major player in the chargino contribution to both 
$b\rightarrow s\gamma$ and $\delta m_b^{SUSY}>0$. As a result of sizable
$A_t$, a significant left-right mixing is induced, leading to the lightest
stop, sbottom and stau decoupled by
up to few hundred GeV from the rest of the sfermions. In fact, 
the experimental lower limit on the stau mass starts affecting the analysis
as soon as $m_0$ gets below 500GeV. 

\begin{figure}[t]
\psfig{figure=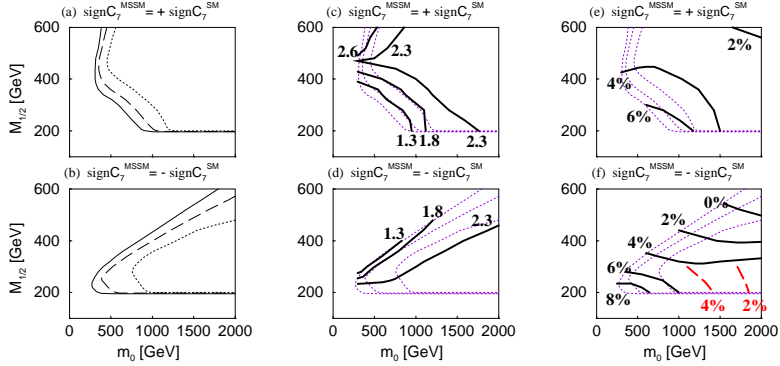,height=1.5in}
\caption{
Model 4c global analysis results for two possible signs of $C_7^{MSSM}$.
In (a) and (b), the best fit contour plots of $\chi^2$ are shown. 
Solid (dashed, dotted) lines correspond to $\chi^2=6\,(3,1)\, per\, 3\,
d.o.f.$. 
Figures (c) and (d), and (e) and (f), show  
the best fit contour plots of $BR(b\rightarrow s\gamma)\times 10^4$
and $\delta m_b^{SUSY}$, respectively, with the $\chi^2$ contour
lines of (a) and (b) in the background for better reference.
}
\end{figure}

To conclude, the results show that the MSSM with large
tan$\beta$ constrained by minimal SO(10) remains an option for physics
beyond the Standard Model. 

This talk is an extraction from a larger project done together with 
Stuart Raby, Marcela Carena and Carlos Wagner, whose help is
greatly appreciated.

\end{document}